# MODELING OF DEFORMATION AND TEXTURE DEVELOPMENT OF COPPER IN A 120° ECAE DIE

R. Arruffat-Massion, L.S. Tóth, J-P. Mathieu
Laboratoire de Physique et Mécanique des Matériaux, ISGMP, Université de Metz, Ile du Saulcy 57045 Metz, France

Keywords : Equal Channel Angular Extrusion, Severe plastic deformation, Crystallographic texture, Copper.

## Abstract

A flow line function is proposed to describe the material deformation in ECAE for a 120° die. This new analytical approach is incorporated into a viscoplastic self-consistent polycrystal code to simulate the texture evolution in Route A of copper and compared to experimental textures as well as to those corresponding to simple shear.

## Introduction

The process of severe plastic deformation is a technique for grain refinement in bulk materials [1]. The most promising is the Equal Channel Angular Extrusion (ECAE), invented in the middle of the 70's and developed in the former Soviet Union by Segal [2-3]. This technique consists in increasing the strain indefinitely by carrying out successive passes in the working device. The die is composed of two channels having the same cross-section and intersecting at an angle $\Phi$.

With the aim of knowing and understanding the evolution of the microstructure, it is essential to analyse the mode of deformation applied during the tests. For a 90° die, Tóth et al. [4] proposed a formulation which made possible to model texture development [5] and strain hardening [6-7]. Some experiments and simulations by finite elements [8-13] have been carried out for a 120° die, but no significant improvement, except for the "fan-model" of Beyerlein [14], was made concerning the strain field in the ECAE die since the model of Segal.

An analytical formula for the equivalent plastic strain was developed by Segal et al. [3] and generalised by Iwahashi et al. [15]:

$$\bar{\varepsilon} = \frac{1}{\sqrt{3}} \left[ 2 \cot\left(\frac{\Phi}{2} + \frac{\Psi}{2}\right) + \Psi \csc\left(\frac{\Phi}{2} + \frac{\Psi}{2}\right) \right]. \tag{1}$$

Here $\Psi$ is the rounding of the die. When the die-corner is not rounded ($\Psi = 0$), this formula simplifies to [3]:

$$\bar{\varepsilon} = \frac{2}{\sqrt{3}} \cot\left(\frac{\Phi}{2}\right). \tag{2}$$

In the configuration of Fig. 1, for $\Phi = 120°$, Eq. (2) gives: $\bar{\varepsilon} = 2/3$. This is the strain that corresponds to the simple shear approximation of the flow field.

The objective of this paper is to propose a new model of deformation for a 120° ECAE die based on a flow line approach. This new flow line function was inspired from the one presented for $\Phi = 90°$ [4] and will be used for the prediction of texture evolution of copper in Route A. The results will be compared to experiments as well as with the prediction of Segal's simple shear model.

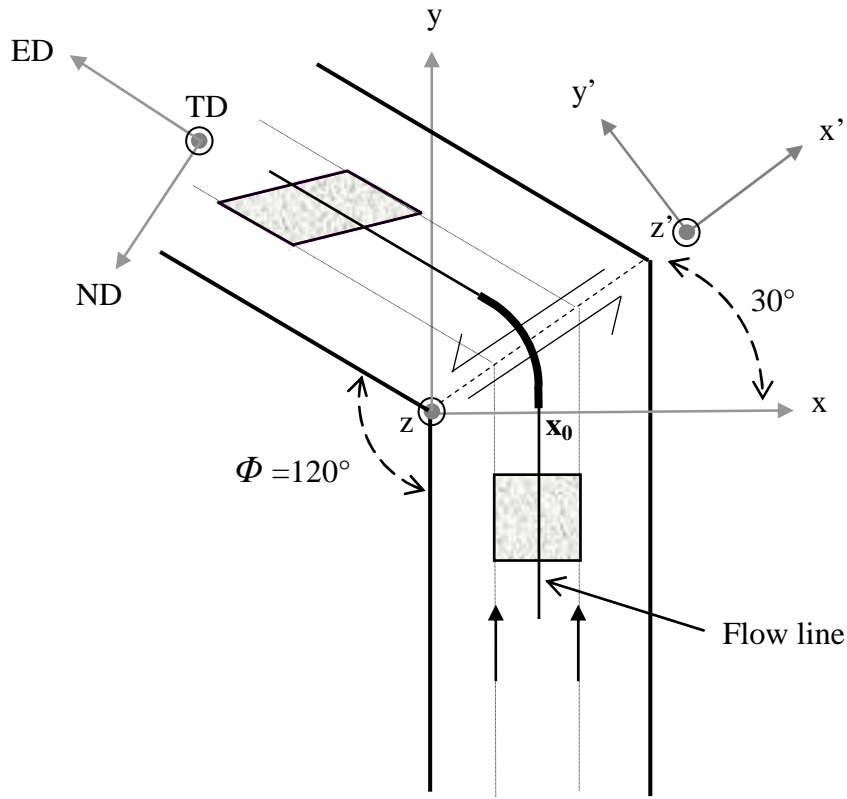

Fig.1 : Description of the ECAE test and the flow field by a flow line.

## **Modeling of deformation**

Two models are considered here to describe the deformation in ECAE: the so-called discontinuous simple shear model, in which the deformation is simple shear localised on the 30° intersection plane of the two channels (see Fig. 1) and a new analytical model using a flow line function.

***The discontinuous simple shear model ('SS'):*** The velocity gradient corresponding to simple shear in the plane of discontinuity (x'-y'-z' reference system in Fig. 1) is given by:

$$L' = \begin{pmatrix} 0 & -\dot{\gamma} & 0 \\ 0 & 0 & 0 \\ 0 & 0 & 0 \end{pmatrix}, \qquad (3)$$

with the rate of shear on the 30° plane, $\dot{\gamma}$, positive. When *L'* is expressed in the x-y-z reference system of Fig. 1, we obtain the following velocity gradient:

$$L = \frac{\dot{\gamma}}{4} \begin{pmatrix} \sqrt{3} & -3 & 0 \\ 1 & -\sqrt{3} & 0 \\ 0 & 0 & 0 \end{pmatrix}. \qquad (4)$$

**L** in Eq. 4 represents a simple proportional loading path, which can be readily incorporated into a polycrystal plasticity code. This simple shear approach is based on Segal's model [3].

***The flow line model ('FL'):*** The following flow function is proposed to approximate the material flow in a 120° ECAE die:

$$\phi = x^n + \left(\frac{1}{2}\right)^n \left[\left(x + \sqrt{3}y\right)^n + \left(-x + \sqrt{3}y\right)^n\right] = x_0^n \qquad (5)$$

where $x_o$ defines the incoming position of the flow line (see Fig. 1) and $n$ is a parameter. It is interesting to note that equation (5) describes a regular hexagon for even values of $n$ whereas the shape of the curve is not the same when $n$ is odd. This characteristic does not obstruct the modeling because only one part of the flow line is used (bold solid line in Fig. 1), which is a sixth of the regular hexagon, so that $x$ and $y$ are positive during the plastic deformation. This is the reason why the reference system was selected in the position shown in Fig. 1. The $n$ parameter describes the possible shapes of the flow lines: for $n = 2$ the line is circular, for higher $n$ values it approximates the flow better within a non-rounded die. For $n \to \infty$, the flow line is constituted by simply two straight lines connected at the 30° plane.

An incompressible velocity field can be defined from the flow function $\phi$ as follows:

$$v_x = \lambda \frac{\partial \phi}{\partial y} = -\frac{v_0 \sqrt{3} D \left[A^{n-1} + B^{n-1}\right]}{x_0^{n-1} E},$$

$$v_y = -\lambda \frac{\partial \phi}{\partial x} = \frac{v_0 \left[C + D\left(A^{n-1} - B^{n-1}\right)\right]}{x_0^{n-1} E}. \qquad (6)$$

where $A = x + \sqrt{3}y$, $B = -x + \sqrt{3}y$, $C = x^{n-1}$, $D = \left(\frac{1}{2}\right)^n$, $E = 1 + 2^{-n} + \left(-\frac{1}{2}\right)^n$.

The λ parameter in Eq. (6) is determined by the incoming velocity of the material $v_o$ as:

$$\lambda = -\frac{v_0}{E n x_0^{n-1}}.$$

To further simplify the above expressions, the following notation is used:

$$F = x^{n-2}, \quad G = x^n + D(A^n + B^n), \quad H = EG^{(n-1)/n}, \quad I = A^{n-1} - B^{n-1},$$
$$J = A^{n-1} + B^{n-1}, \quad K = A^{n-2} + B^{n-2}, \quad L = A^{n-2} - B^{n-2}.$$

Using the above notation, the flow line function leads to the following velocity gradient:

$$L_{xx} = \frac{3v_0 D(n-1)}{E}\left\{-\frac{K}{x_0^{n-1}} + \frac{J(C+DI)}{x_0^{2n-1}}\right\},$$

$$L_{xy} = \frac{3v_0 D(n-1)}{E}\left\{-\frac{K}{x_0^{n-1}} + \frac{DJ^2}{x_0^{2n-1}}\right\},$$

$$L_{yx} = \frac{v_0(n-1)}{E}\left\{\frac{F+DK}{x_0^{n-1}} - \frac{[C+DI]^2}{x_0^{2n-1}}\right\}$$

$$L_{yy} = -L_{xx}, \quad L_{yz} = L_{zx} = L_{zy} = L_{zz} = 0. \tag{7}$$

The strain rate tensor is the symmetrical part of **L**:

$$\dot{\varepsilon}_{xy} = \frac{v_0(n-1)}{2H}\left\{3D\left(-K + \frac{DJ^2}{G}\right) + F + DK - \frac{(C+DJ)^2}{G}\right\},$$

$$\dot{\varepsilon}_{xx} = L_{xx}, \quad \dot{\varepsilon}_{yx} = \dot{\varepsilon}_{xy}, \quad \dot{\varepsilon}_{yy} = L_{yy}, \quad \dot{\varepsilon}_{yz} = \dot{\varepsilon}_{zx} = \dot{\varepsilon}_{zy} = \dot{\varepsilon}_{zz} = 0. \tag{8}$$

The equivalent strain rate in the sense of von Mises is:

$$\dot{\bar{\varepsilon}} = \frac{v_0 2(n-1)D}{H}\left\{\left[-L + \frac{(C+DI)J}{G}\right]^2 + \left[\frac{\sqrt{3}}{2}\left(-K + \frac{D}{G}J^2\right) + \frac{\sqrt{3}}{6D}\left(F + DK - \frac{(C+DI)^2}{G}\right)\right]^2\right\}^{1/2}. \tag{9}$$

With the difference of the model for a 90° die proposed by Tóth et al. [4], an analytical integration was not possible along a flow line to obtain the total accumulated equivalent strain in one pass. A numerical integration has been carried out; the results are presented in figure 2. Comparisons are possible with the simple shear model (Eq. (1)) for $\Psi=60°$ and $\Psi=0°$. For $\Psi=60°$, Eq. (1) gives $\pi/(3\sqrt{3})$ which is the minimal value of the equivalent strain (maximum rounding of the die). The same result is obtained from the modeling by the present flow line approach with $n = 2$. When $n$ tends towards infinity, the total accumulated equivalent strain approaches the value of 2/3. This is the Segal's value which belongs to $\Psi=0°$ in Eq. 1. The equivalent strain in one pass continuously increases between these limits, as it is displayed in Fig. 2.

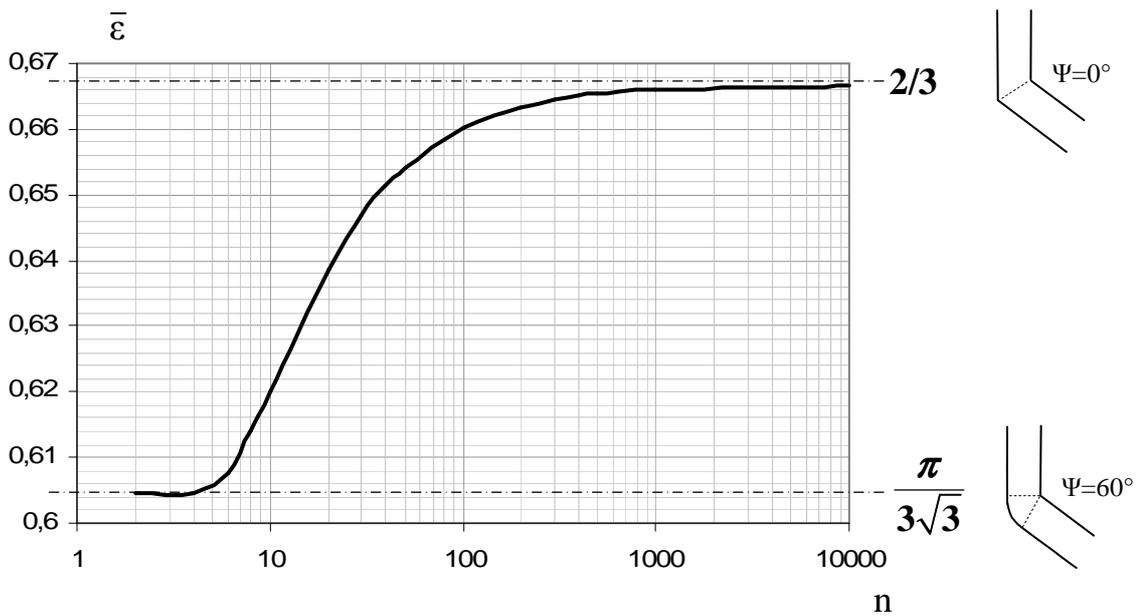

Fig.2 : The von Mises accumulated strain in one pass as a function of the *n* exponent of the flow line function.

**<u>Experimental procedure and results</u>**

The experimental results used here were published by Schafler et al. [9-10]. Copper samples with 15×15 mm square cross-section and 35 mm length were machined from annealed bars. The ECAE experiments were carried out at a die set with rectangular shape of the extrusion channels without any rounding of the corner region ($\Psi = 0$, see Fig. 1) at room temperature. The specimens were extruded in Route A, i.e. the samples were not rotated between successive passes around their long axis. Texture measurements were carried out on the ED-ND plane of the sample (Fig. 1) by X-Ray diffraction (Cu K$_\alpha$ radiation) on four diffraction planes $\{111\},\{200\},\{220\}$ and $\{311\}$. A detailed analysis of the texture was carried out using the method of orientation distribution functions [16]. Crystals orientations are defined in terms of Euler angles $\phi_1,\phi,\phi_2$ in Bunge's notation with respect to the ED, ND and TD axes which are the $x_1, x_2, x_3$ axes for the Euler angles. In order to save space, only the $\phi_2 = 0°$ and the $\phi_2 = 45°$ sections of the ODFs are shown in Fig. 3. A key figure showing the ideal positions of simple shear is also presented in Fig. 3. The ODFs are presented up to three passes. Their examination reveals that the general strength of the texture does not change much but there are significant intensity variations in the individual components. The C$_E$ component seems to disappear in the third pass, in reality it persists, just its intensity decreases below 2 (to 1.6).

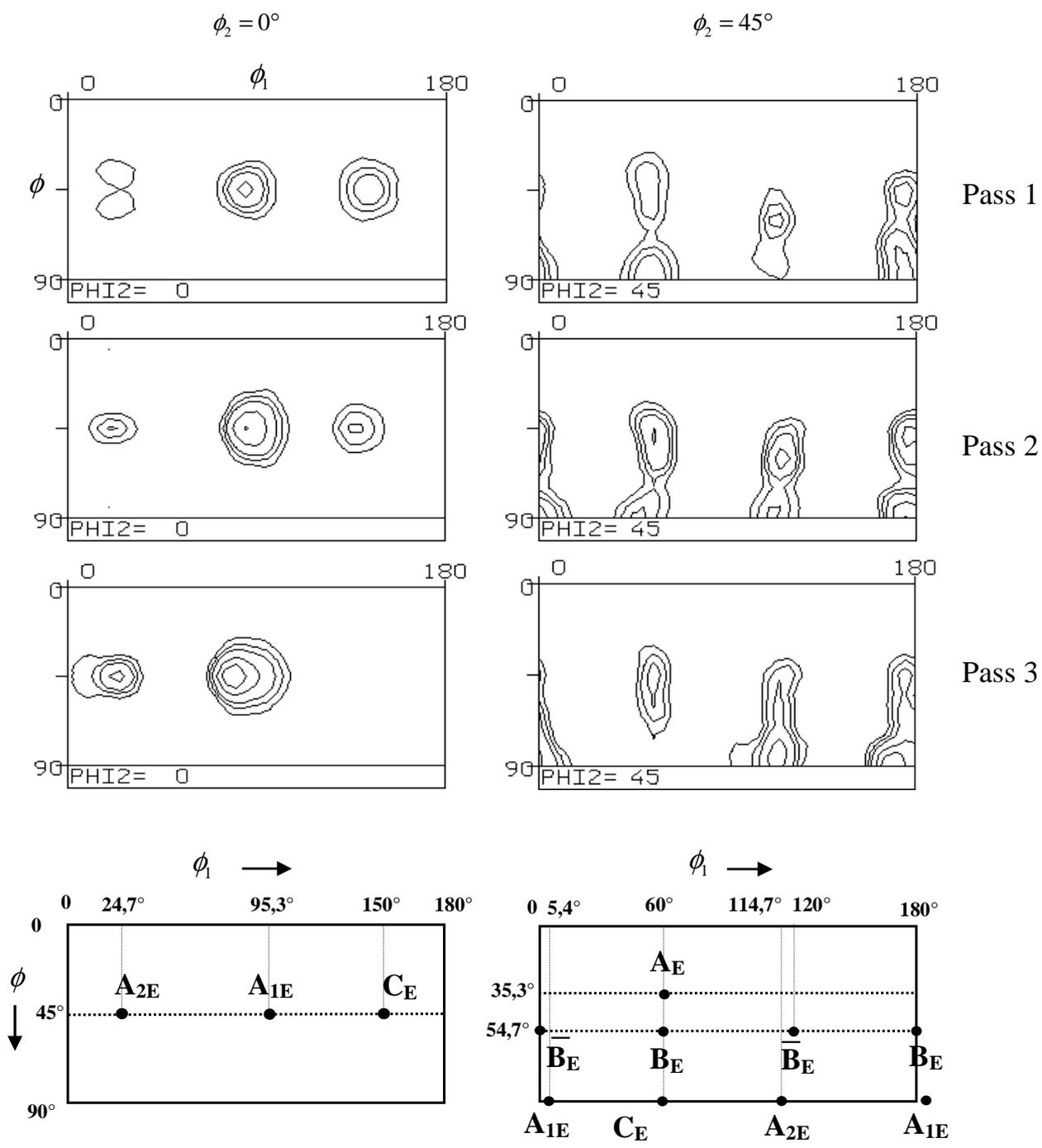

Fig.3: Experimental textures obtained in Copper for Route A.
(isovalues: 2, 3, 4, 6, 8, 10, 12, 16, 20, 24.)

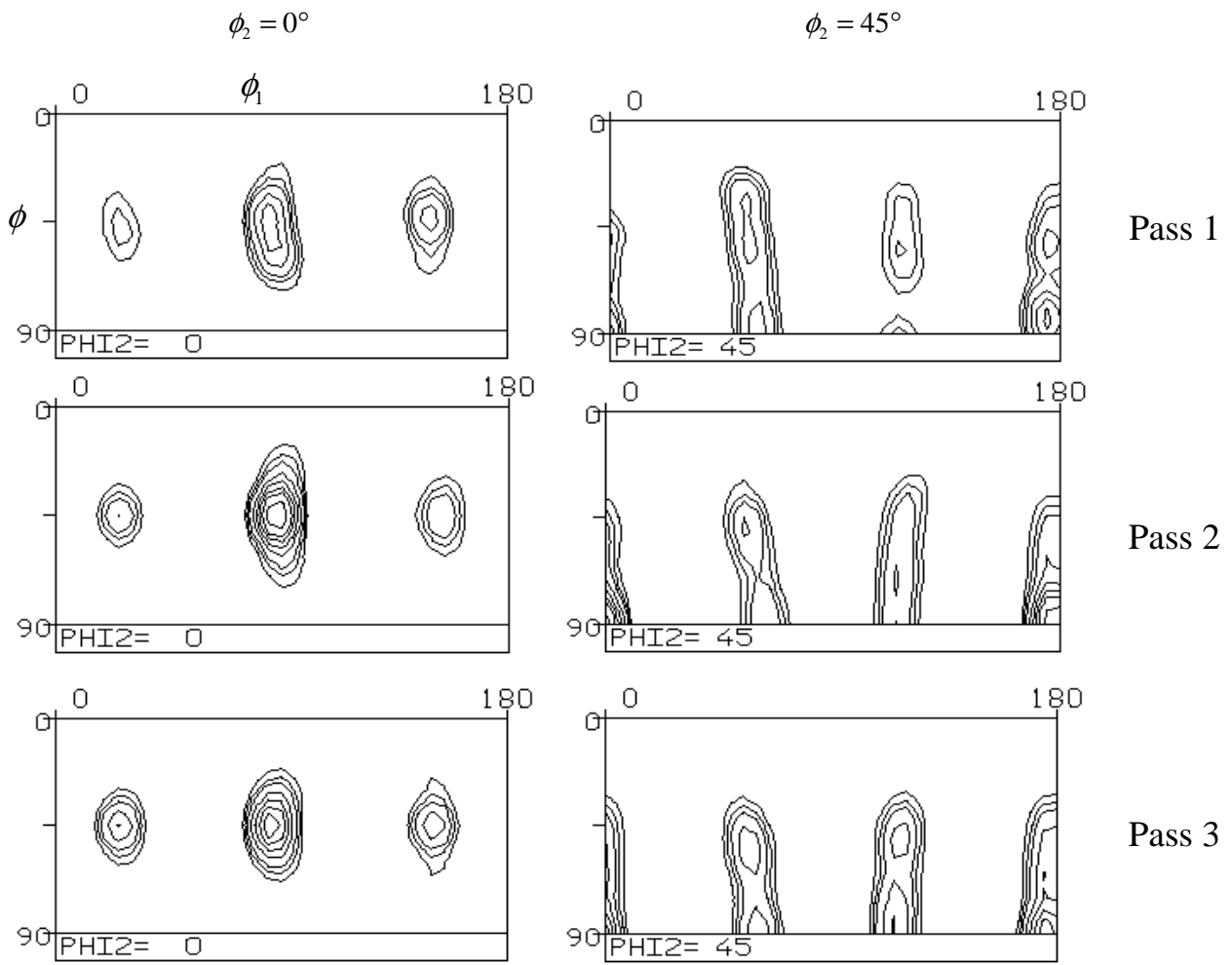

Fig.4: Simulated textures obtained with discontinuous shear model and self consistent in Copper for Route A. (isovalues: 2, 3, 4, 6, 8, 10, 12, 16, 20, 24.)

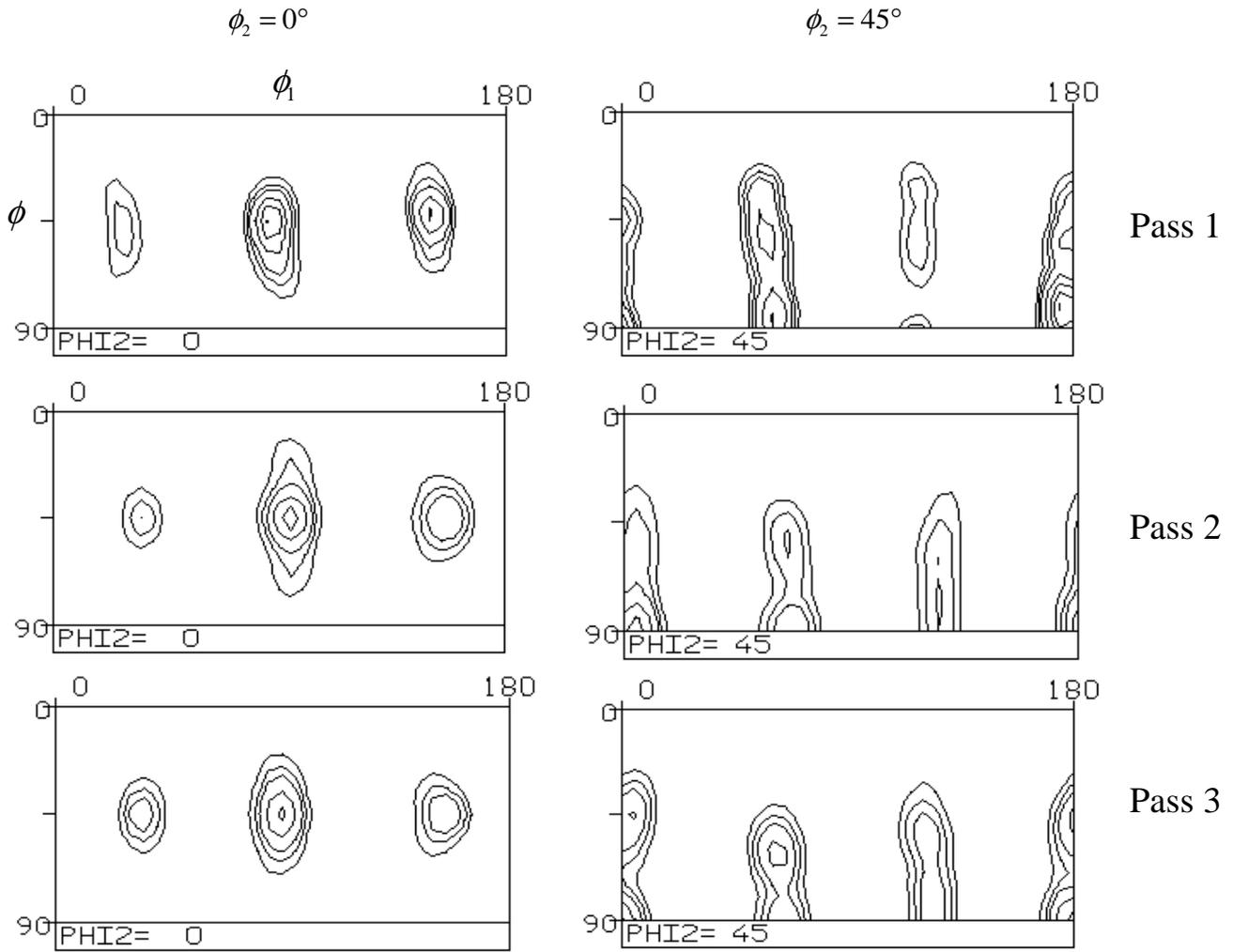

Fig.5: Simulated textures obtained with flow line and self consistent in Copper for Route A with *n*=12,4,4. (isovalues: 2, 3, 4, 6, 8, 10, 12, 16, 20, 24.)

**Simulations**

The two above described approaches were introduced into the Self-Consistent (SC) polycrystal model (the finite-element tuned SC model was employed in its isotropic version). The strain rate sensitivity index of crystallographic slip was chosen to be 0.05. The 12 {111}<110> slip systems of f.c.c. crystal were employed with self and latent hardening [4]. It is to be noted that the effect of hardening on the evolution of texture was found to be very small so it was neglected. Simulations have been carried out using the velocity gradient (Eq. 7) in an incremental way. The initial texture was represented by 500 randomly oriented grains. For the second and the third passes, the "texture corrected" technique [4] was used: that is, the measured texture was discretized (3000 grains) and introduced into the polycrystal code as input texture for the next pass.

In the discontinuous simple shear model, a constant velocity gradient was used (eq. (4)) in an incremental manner up to the shear value of $\gamma = 2/(3\sqrt{3})$. The simulated textures obtained with this modeling are presented in Fig. 4.

In the flow line model, the texture was placed on a flow line at $x = x_0$, $y = 0$ and subjected to the strain field defined by the velocity gradient in Eq. (7). That flow line was selected which passes in the middle of the die, i.e. at $x_0 = d/2$, where $d$ is the size of the channel. The physical displacement of the material element was followed using the local velocity (Eq. (6)) and the expression of the flow function (Eq. (5)). For the choice of the $n$ parameter of the flow field, that value of $n$ was used which led to the best agreement between the simulated textures and the experiments. In this way, the parameter $n$ was found to be 12 for the first pass and 4 for the two following passes. The results obtained with the flow line approach are displayed in Fig. 5.

**Discussion**

The main purpose of the present paper is to compare the predicting capacity of the two models, i.e. the simple shear and the flow line approaches. As can be seen from a comparison of the experimental textures (Fig. 3) to that of the predicted ones using the SS (Fig. 4) and FL models (Fig. 5), one can see that the main features are well predicted by both approaches. There were more deviations for textures predicted for a 90° die when the above models were applied [4]. The reason for the better agreement between the two models in the present 120° die could be that the total strain is significantly less, so deviations are expected to be less developed. Also, the rotation of the texture due to the repositioning of the sample into the die in a subsequent pass with respect to its previous position is smaller in a 120° die (it is only 60° in the $\phi_1$ angle in orientation space with respect to 90° in a 90° die) which should lead to less deviations between the respective development of the textures. This difference means otherwise that the strain path deviation is less between subsequent passes in a 120° die with respect to a 90° die.

When the predicted intensities of the textures are compared to the experimental ones, however, one can see that the FL model predicts much lower intensities, nearly the same as the experimental ones. This is a very positive result which can justify the use of such a modeling even if the mathematics seems to be much more complicated than in the SS model.

One could also examine the relative shifts of the components from their ideal positions in the present textures, like for the 90° case where they were reported to be significant and systematic [4]. In the present experiments, however, 'global' textures were measured (on the whole surface of the side plane ED-ND which makes difficult to evaluate the rotation exactly. Namely, it has been shown in [17] that a relatively strong texture gradient can develop perpendicular to the flow direction in the die, concerning especially the 'tilts' of the components from their ideal positions. As the measured texture is detected irrespectively from this gradient, such analysis was not carried out in the present work and could be the subject of future investigations.

## Conclusion

In the present work, the strain field for a 120° die in ECAE has been investigated with the help of a new flow line function and was compared to Segal's simple approach. Both models were applied to predict texture evolution in copper subjected to Route A. It has been shown that the new flow line model predicts textures in better agreement with the experiments, especially concerning the intensities of the components.

## Acknowledgement

The Authors acknowledge gratefully the experimental ODF data for the present ECAE Route A copper made available by E. Schafler and I. Kopacz (University of Vienna).